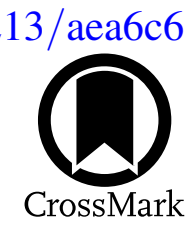

# Measuring the Expansion of Solar Magnetic Fields with Multiline Inversions from Sunrise III

Juan Carlos Trelles Arjona[1,2], María Jesús Martínez González[1,2], Christoph Kuckein[1,2,3], Sergio Javier González Manrique[1,2], Manuel Collados[1,2], Sami K. Solanki[4], Andreas Lagg[4], Achim Gandorfer[4], Jose Carlos del Toro Iniesta[3,5], Yukio Katsukawa[6,7,8], Pietro Bernasconi[9], Thomas Berkefeld[10], Alex Feller[4], Tino L. Riethmüller[4], Alberto Álvarez-Herrero[3,11], Masahito Kubo[6,8], H. N. Smitha[4], David Orozco Suárez[3,5], Bianca Grauf[4], Michael Carpenter[9], Alexander Bell[10], Valentín Martínez Pillet[1,3,12], Francisco Javier Bailén[3,5], Julian Blanco Rodríguez[3,13], Juan Sebastián Castellanos Durán[4], Edvarda Harnes[4], Johannes Hölken[4], Francisco A. Iglesias[4,14], Ryohtaroh T. Ishikawa[15], Yusuke Kawabata[6], Takuma Matsumoto[16], Takayoshi Oba[4,17], Azaymi L. Siu-Tapia[3,5], Hanna Strecker[3,5], Dušan Vukadinovic[4,18], Hirohisa Hara[6], and Toshifumi Shimizu[19,20]

[1] Instituto de Astrofísica de Canarias, Vía Láctea s/n, E-38205 La Laguna, Spain; juan.trelles@iac.es
[2] Departamento de Astrofísica, Universidad de La Laguna, E-38206 La Laguna, Spain
[3] Spanish Space Solar Physics Consortium, Spain
[4] Max-Planck-Institut für Sonnensystemforschung, Justus-von-Liebig-Weg 3, 37077 Göttingen, Germany
[5] Instituto de Astrofísica de Andalucía, CSIC, Glorieta de la Astronomía s/n, 18008 Granada, Spain
[6] National Astronomical Observatory of Japan, 2-21-1 Osawa, Mitaka, Tokyo 181-8588, Japan
[7] Department of Astronomy, The University of Tokyo, 7-3-1, Hongo, Bunkyo-ku, Tokyo 113-0033, Japan
[8] Department of Astronomical Science, The Graduate University for Advanced Studies (SOKENDAI), 2-21-1 Osawa, Mitaka, Tokyo 181-8588, Japan
[9] Johns Hopkins University Applied Physics Laboratory, 11100, Johns Hopkins Road, Laurel, MD, USA
[10] Institut für Sonnenphysik (KIS), Georges-Köhler-Allee 401a, 79110 Freiburg, Germany
[11] Instituto Nacional de Técnica Aeroespacial (INTA), Carretera de Ajalvir, kilómetro 4, E-28850 Torrejón de Ardoz, Spain
[12] Universidad de La Laguna, E-38205 La Laguna, Spain
[13] Universitat de Valencia, Catedrático José Beltrán 2, E-46980 Paterna, Spain
[14] Grupo de Estudios en Heliofísica de Mendoza, CONICET, Universidad de Mendoza, Boulogne sur Mer 683, 5500 Mendoza, Argentina
[15] National Institute for Fusion Science, 322-6 Oroshi-cho, Toki City 509-5292, Japan
[16] Centre for Integrated Data Science, Institute for Space-Earth Environmental Research, Nagoya University, Furocho, Chikusa-ku, Nagoya, Aichi 464-8601, Japan
[17] Advanced Research Center for Space Science and Technology, Institute of Science and Engineering, Kanazawa University, Kakuma-machi, Kanazawa, Ishikawa 920-1192, Japan
[18] Institut für Physik, Universität Graz, Universitätsplatz 5, 8010 Graz, Austria
[19] Department of Earth and Planetary Science, The University of Tokyo, 7-3-1, Hongo, Bunkyo-ku, Tokyo 113-0033, Japan
[20] Institute of Space and Astronautical Science, Japan Aerospace Exploration Agency, 3-1-1, Yoshinodai, Chuo-ku, Sagamihara, Kanagawa 252-5210, Japan

## Abstract

Magnetic fields in the solar atmosphere are expected to expand with height, forming magnetic canopies that couple the photosphere and chromosphere. Except for sunspots, direct measurements of this expansion remain scarce because they require simultaneous magnetic diagnostics over a broad range of atmospheric heights. We analyze four consecutive high-resolution spectropolarimetric raster scans of an emerging flux region observed with the Sunrise III/SCIP instrument. Simultaneous inversions of five photospheric Fe I lines and two chromospheric Ca II lines provide a quasi-continuous reconstruction of the magnetic field from the deep photosphere to the lower chromosphere. We apply a constant-flux method that follows the same magnetic flux through different atmospheric layers to quantify magnetic expansion. All three pores analyzed exhibit an overall increase in magnetic area with height, approximately doubling their cross section between the lower photosphere and the highest atmospheric layers sampled by the inversions. The expansion remains broadly stable throughout the 48 minute observing sequence. We further characterize the expansion with a quadratic parameterization and investigate its relation to pore properties. The three pores exhibit similar core magnetic field strengths but different expansion behaviors, indicating that the magnetic core strength alone does not determine the expansion. However, the present sample is too small to establish whether these differences are related to the thermodynamic or magnetic properties of the pores and their surroundings. These results demonstrate the potential of simultaneous multiline inversions for investigating the three-dimensional magnetic topology of solar magnetic structures and provide the first direct measurements of the height-dependent magnetic expansion of solar pores.

## 1. Introduction

The magnetic flux concentrations within magnetic flux emerging regions in the solar lower atmosphere evolve toward predominantly vertical configurations. These vertical magnetic concentrations expand with height in response to the rapid decrease of the external gas pressure. This expansion causes the magnetic field to extend laterally beyond the photospheric concentration and the field lines at its periphery to become progressively more inclined (H. C. Spruit 1976; S. K. Solanki 1993). At sufficiently high atmospheric layers, this geometry can lead to the formation of a magnetic canopy, understood here as a spatially extended magnetic structure overlying the surrounding atmosphere and containing a significant inclined or horizontal magnetic-field component (S. K. Solanki 1993).

Classical magnetic flux tube models predict the formation of magnetic canopies in the upper photosphere and lower chromosphere (H. C. Spruit 1976). Modern radiative magnetohydrodynamic simulations reproduce this behavior (e.g., S. K. Solanki & O. Steiner 1990; for a review, see O. Steiner 2010), showing that magnetic fields expand laterally, become progressively weaker, and develop increasingly inclined configurations toward higher atmospheric layers.

Observational evidence for magnetic canopies has been reported in several studies. Using Sunrise/IMaX observations (P. Barthol et al. 2011; V. Martínez Pillet et al. 2011), M. J. Martínez González et al. (2012) inferred the presence of low-lying magnetic canopies surrounding quiet Sun magnetic elements from the spatial distribution of Stokes $V$ asymmetries, suggesting the existence of strong line of sight (LoS) gradients associated with expanding magnetic fields. Likewise, A. Pietarila et al. (2010) used center to limb observations to investigate the expansion of magnetic flux concentrations, building on the theoretical interpretation developed by U. Grossmann-Doerth et al. (1988) and S. K. Solanki (1989). More recently, C. Kuckein (2019) studied the height variation of the magnetic field and plasma flows in isolated bright points using spectropolarimetric observations spanning the photosphere and chromosphere, while D. Buehler et al. (2015) combined high-resolution spectropolarimetric observations with numerical simulations to investigate the vertical magnetic structure of magnetic flux concentrations. R. Morosin et al. (2020) reconstructed the magnetic-field stratification of a plage region using multiline spectropolarimetric observations and found that photospheric magnetic concentrations expand horizontally toward the chromosphere, where the field becomes increasingly space filling and forms a magnetic canopy. However, these studies were largely limited by the atmospheric layers sampled or relied on indirect diagnostics of the magnetic topology.

Directly characterizing the height-dependent expansion of magnetic structures requires high-sensitivity spectropolarimetric observations capable of sampling multiple atmospheric layers simultaneously with sufficient spatial and spectral resolution. Magnetic expansion is a fundamental property for understanding and constraining the formation of magnetic canopies. The advent of multiline instruments such as the Spectropolarimetric and Chromospheric Imager (SCIP; Y. Katsukawa et al. 2026) aboard the Sunrise III balloon observatory (A. Korpi-Lagg et al. 2025) now enables this by combining several photospheric and chromospheric spectral lines observed simultaneously at the spatial resolution of a 1 m telescope.

In this letter, we exploit the novel high-spatial-resolution, multiline spectropolarimetric data to perform simultaneous inversions of five photospheric Fe I lines together with the chromospheric Ca II 8498 and 8542 Å lines recorded by SCIP. These inversions constrain the magnetic-field stratification from the deep photosphere to the lower chromosphere and demonstrate the potential of this approach for studying the three-dimensional magnetic structure of the solar atmosphere (e.g., T. L. Riethmüller & S. K. Solanki 2019; J. C. Trelles Arjona et al. 2021; J. C. Trelles Arjona 2025; J. Hölken et al. 2026). Building on these atmospheric models, we quantify the expansion of solar magnetic structures as a function of optical depth, i.e., height, and time, and investigate whether the observed differences in expansion are related to the physical properties of the pores and their surroundings.

## 2. Observations

The observations analyzed in this work were obtained on 2024 July 15 with the Sunrise III balloon-borne solar observatory during its third science flight (A. Korpi-Lagg et al. 2025; S. K. Solanki et al. 2026). The target was an emerging flux region located in NOAA active region 13753, observed at heliocentric positions ranging from $\mu \approx 0.96$ at the beginning of the time series to $\mu \approx 0.95$ at its end. The region contains pores and sunspots of opposite magnetic polarity in the photosphere together with an arch filament system and a filament in the chromosphere. Other analyses of this emerging flux region have focused on Ellerman bombs (Y. Kawabata et al. 2026) and on the associated arch filament system (S. González Manrique et al. 2026).

We make use of spectropolarimetric observations acquired with the SCIP instrument (Y. Katsukawa et al. 2026), which consists of two simultaneous spectral channels. In this work, we use only the 850 nm channel, which recorded the full Stokes vector over the wavelength range 8466–8547 Å in several photospheric and chromospheric spectral lines. Figure 1 presents an overview of the observations, showing monochromatic intensity and polarization maps at representative wavelength positions within the Fe I 8468 Å and Ca II 8542 Å profiles, together with the colored circles marking the three pores analyzed in this letter. The wavelength offsets, $\Delta\lambda$, quoted in Figure 1 are measured relative to the reference wavelengths $\lambda_0 = 8468.41$ Å for Fe I and $\lambda_0 = 8542.09$ Å for Ca II. While only the Fe I 8468 Å and Ca II 8542 Å lines are shown here to illustrate the observations, the complete Stokes spectra of seven spectral lines spanning the photosphere and chromosphere are used simultaneously as input to the inversions.

SCIP acquired repeated raster scans covering an FoV of approximately $58'' \times 58''$, sampled with 621 slit positions and a spectral sampling of 39.5 mÅ pixel$^{-1}$. The integration time was 1 s per slit position, yielding a polarimetric sensitivity of approximately $1.1 \times 10^{-3} I_c$ in Stokes $Q$, $U$, and $V$, estimated from the standard deviation of continuum wavelength samples around the Fe I 8468 Å line. Each raster requires approximately 12 minutes to complete. We analyze four consecutive raster maps recorded between 11:40:26 and 12:28:08 UT, covering the intervals 11:40:26–11:52:21 UT (Map 1), 11:52:22–12:04:17 UT (Map 2), 12:04:18–12:16:12 UT (Map 3), and 12:16:14–12:28:08 UT (Map 4). The four consecutive raster maps allow us to investigate both the expansion of solar magnetic structures through the atmosphere along the LoS and their temporal evolution over nearly 48 minutes.

## 3. Methods

### 3.1. Spectropolarimetric Inversions with DeSIRe

The three-dimensional magnetic structure analyzed in this work was inferred from simultaneous multiline inversions performed with the non–local thermodynamic equilibrium (NLTE) DeSIRe inversion code (B. Ruiz Cobo et al. 2022). The inversions included the photospheric Fe I lines at 8468 Å ($g_{\rm eff} = 2.50$), 8471 Å ($g_{\rm eff} = 1.50$), 8497 Å ($g_{\rm eff} = 1.50$), 8514 Å ($g_{\rm eff} = 1.83$), and 8515 Å ($g_{\rm eff} = 0.75$), together with the chromospheric Ca II 8498 Å ($g_{\rm eff} = 1.07$) and 8542 Å ($g_{\rm eff} = 1.10$) lines, all of them exhibiting significant Zeeman sensitivity. The simultaneous inversion of five photospheric Fe I lines and two chromospheric Ca II lines provides a well-constrained stratification of the magnetic field from the deep photosphere to the midchromosphere (C. Quintero Noda et al. 2017).

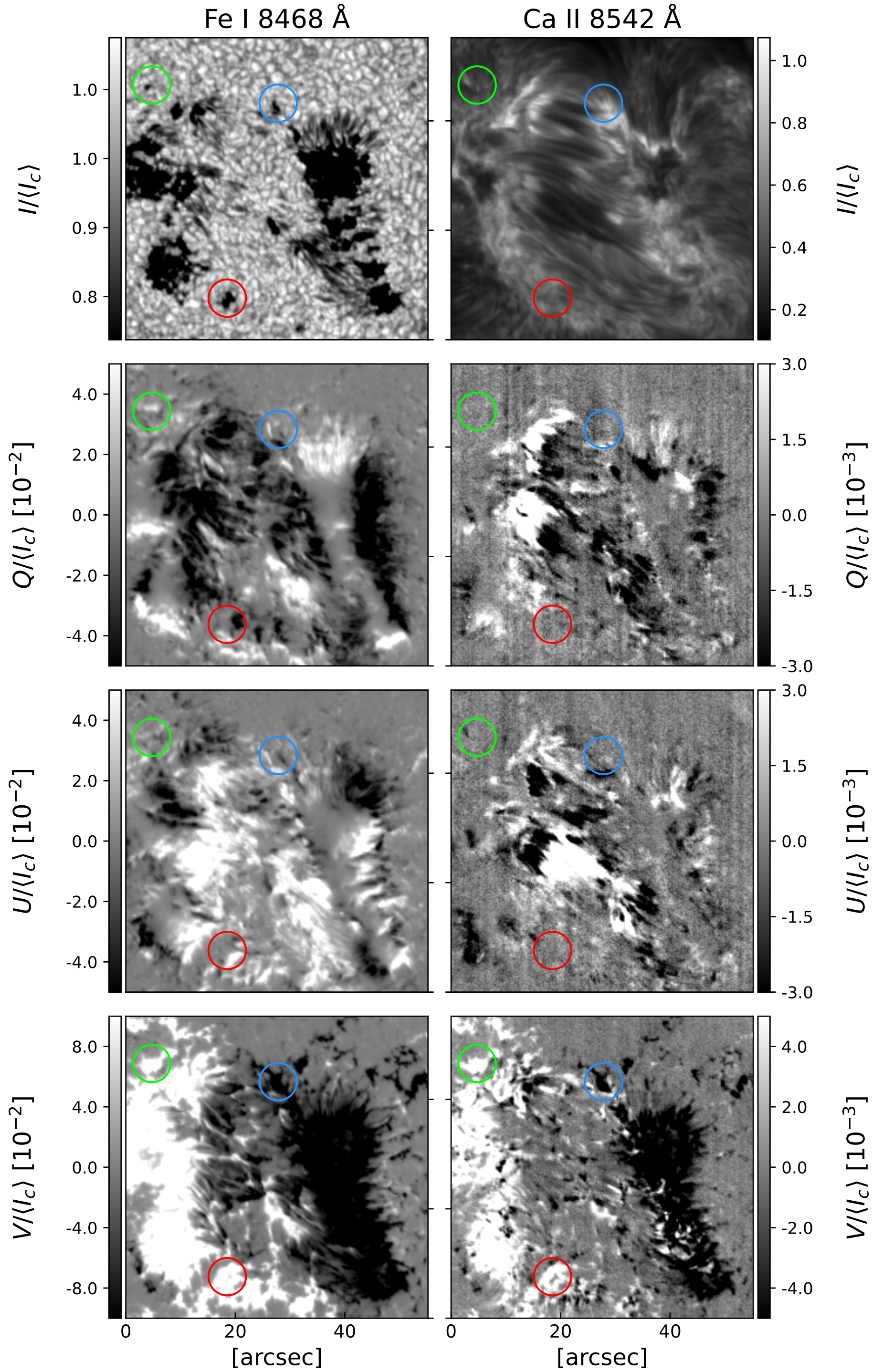

**Figure 1.** Overview of the Sunrise III/SCIP observations analyzed in this letter. The field of view (FoV) covers approximately $58'' \times 58''$. The left column shows, from top to bottom, the Fe I 8468 Å continuum intensity map ($\Delta\lambda = -0.551$ Å), representative of the photosphere, followed by monochromatic Stokes $Q$, $U$, and $V$ maps obtained at $\Delta\lambda = -0.117$ Å. The right column displays, from top to bottom, the Ca II 8542 Å line core intensity map ($\Delta\lambda = -0.008$ Å), sampling the chromosphere, followed by monochromatic Stokes $Q$ and $U$ maps obtained at $\Delta\lambda = -0.087$ Å and a Stokes $V$ map obtained at $\Delta\lambda = -0.403$ Å. All Stokes maps are expressed in units of the mean continuum intensity, $\langle I_c \rangle$. The display ranges are intentionally saturated to enhance the visibility of small-scale structures, with the color bar limits indicating the adopted saturation levels. The red, green, and blue circles in all panels indicate the locations of Pores 1, 2, and 3, respectively. The $x$-axis corresponds to the slit direction, whereas the $y$-axis corresponds to the scan direction.

DeSIRe iteratively retrieves the atmospheric stratification by solving the polarized radiative transfer equation to reproduce the observed Stokes profiles, treating the photospheric Fe I lines under the local thermodynamic equilibrium (LTE) approximation, while the atomic level populations of the chromospheric Ca II lines are computed by solving the statistical equilibrium equations under NLTE conditions, assuming unpolarized atomic levels. Consequently, the polarization of the spectral lines is entirely attributed to the Zeeman effect. This approximation neglects scattering polarization and the Hanle effect, which can affect the inferred transverse magnetic field, particularly in weak field regions. However, the constant-flux analysis presented below relies on the LoS magnetic-field component, primarily constrained by Stokes $V$, for which Zeeman NLTE inversions of Ca II 8542 Å line have been shown to provide reliable estimates (R. Centeno et al. 2021). The inversion setup allowed for height-dependent variations of temperature, LoS velocity, microturbulent velocity, magnetic-field strength, inclination, and azimuth through multiple nodes in optical depth. The atmospheric stratification is reconstructed by smoothly interpolating between adjacent nodes, allowing LoS gradients to be recovered while preserving physically smooth atmospheric stratifications.

The inversion strategy consisted of four successive inversion cycles. In each cycle, the atmospheric model obtained in the previous cycle was refined by increasing the number of equidistant nodes while maintaining a single atmospheric component and equal weights for the four Stokes parameters. In the final cycle, the model atmosphere was parameterized using up to 11 nodes for temperature, 7 for the magnetic-field strength and LoS velocity, and 3 for the microturbulent velocity, magnetic-field inclination, and magnetic-field azimuth.

The synthetic spectra were spectrally degraded using the measured instrumental spectral point-spread function (PSF). The spectral PSF and the wavelength-independent veil intensity ($I_v$) were estimated by comparing the average quiet Sun Stokes $I$ profile, obtained from dedicated SCIP quiet Sun observations, with the solar Fourier Transform Spectrometer (FTS) atlas (L. Wallace & W. Livingston 2003). The best agreement was obtained for a Gaussian PSF with an FWHM of 78.59 mÅ and a veil intensity corresponding to 3.4% of the quiet Sun continuum intensity.

The observed Stokes profiles were corrected for the wavelength-independent, unpolarized veil contribution according to

$$I_{\rm corr} = \frac{I - \alpha I_{\rm c}(x, y)}{1 - \alpha}, \qquad (1)$$

where $I_{\rm c}(x, y)$ is the local continuum intensity at each spatial pixel, estimated by averaging continuum wavelength samples, and $\alpha = 0.034$ is the fractional veil contribution. Since the veil is assumed to be unpolarized, no veil signal was subtracted from Stokes $Q$, $U$, and $V$. These profiles were corrected only for the corresponding dilution of the polarization signal, such that $Q_{\rm corr} = Q/(1 - \alpha)$, $U_{\rm corr} = U/(1 - \alpha)$, and $V_{\rm corr} = V/(1 - \alpha)$. The four corrected Stokes profiles, $I_{\rm corr}$, $Q_{\rm corr}$, $U_{\rm corr}$, and $V_{\rm corr}$, were used as input to the multiline inversions.

The inversions were initialized from three atmospheric models based on the semiempirical FAL-C quiet Sun atmosphere (J. M. Fontenla et al. 1993). The initial magnetic-field strength, inclination, and azimuth were randomly perturbed within representative ranges typical of quiet Sun, penumbral, and umbral magnetic fields in order to reduce the probability of converging toward local minima of the merit function. For each pixel, the final solution was selected as the one yielding the lowest $\chi^2$, which quantifies the goodness of fit between the observed and synthesized Stokes profiles. Figure 2 shows representative observed and best-fit Stokes profiles for two pixels located in Pore 1 (left column) and Pore 2 (right column), marked by the red and green circles in Figure 1, respectively, and sampling different magnetic-field strengths. The agreement between the observed and synthetic profiles is very good. Throughout this work, $\tau$ denotes the continuum optical depth at 500 nm, with $\log \tau = 0$ approximately corresponding to the formation height of the 500 nm continuum and increasingly negative values representing progressively higher atmospheric layers. To assess the height range over which the inferred magnetic stratification is constrained by the observations, we additionally calculated magnetic response functions for representative pore atmospheres. These calculations confirm that the Ca II 8498 and 8542 Å lines retain sensitivity to the magnetic-field strength and inclination around $\log \tau = -4.5$, consistent with previous studies of the Ca II infrared triplet (C. Quintero Noda et al. 2017; A. G. M. Pietrow et al. 2020). We therefore adopt $\log \tau = -4.5$ as the upper boundary of our quantitative analysis.

Figure 3 illustrates the magnetic-field strength, inclination, and azimuth retrieved from the simultaneous multiline inversions covering the full FoV observed by SCIP at different optical depth layers. As expected from the gravity-imposed pressure stratification, the magnetic-field strength decreases progressively with height while the magnetic structures become increasingly extended as a consequence of magnetic flux conservation.

For visualization purposes only, we first determine, at each optical depth layer, a mask for which the integrated longitudinal magnetic flux across all pixels in the FoV approximately matches that of the reference layer at $\log \tau = -0.5$. The mean magnetic-field strength within this mask, $B_{\rm scale}$, is then used to normalize the corresponding magnetic-field map. This normalization emphasizes the spatial expansion of the magnetic structures while allowing a direct comparison of their morphology over the full height range. The normalization is applied exclusively for the visualization shown in Figure 3 and is not used in any of the quantitative analysis presented below. The inclination maps show the progressive lateral expansion of the magnetic field associated with the pores, causing their predominantly vertical fields to occupy an increasingly larger fraction of the FoV toward the upper atmospheric layers. The azimuth maps preserve a coherent large-scale spatial organization throughout the sampled atmosphere.

### 3.2. Constant-flux Expansion Method

To quantify the expansion along the LoS of the magnetic structures, we employed a constant-flux approach that follows the same magnetic flux through different atmospheric layers. At each layer, the method identifies the connected magnetic structure enclosing the reference flux and measures the evolution of its cross-sectional area.

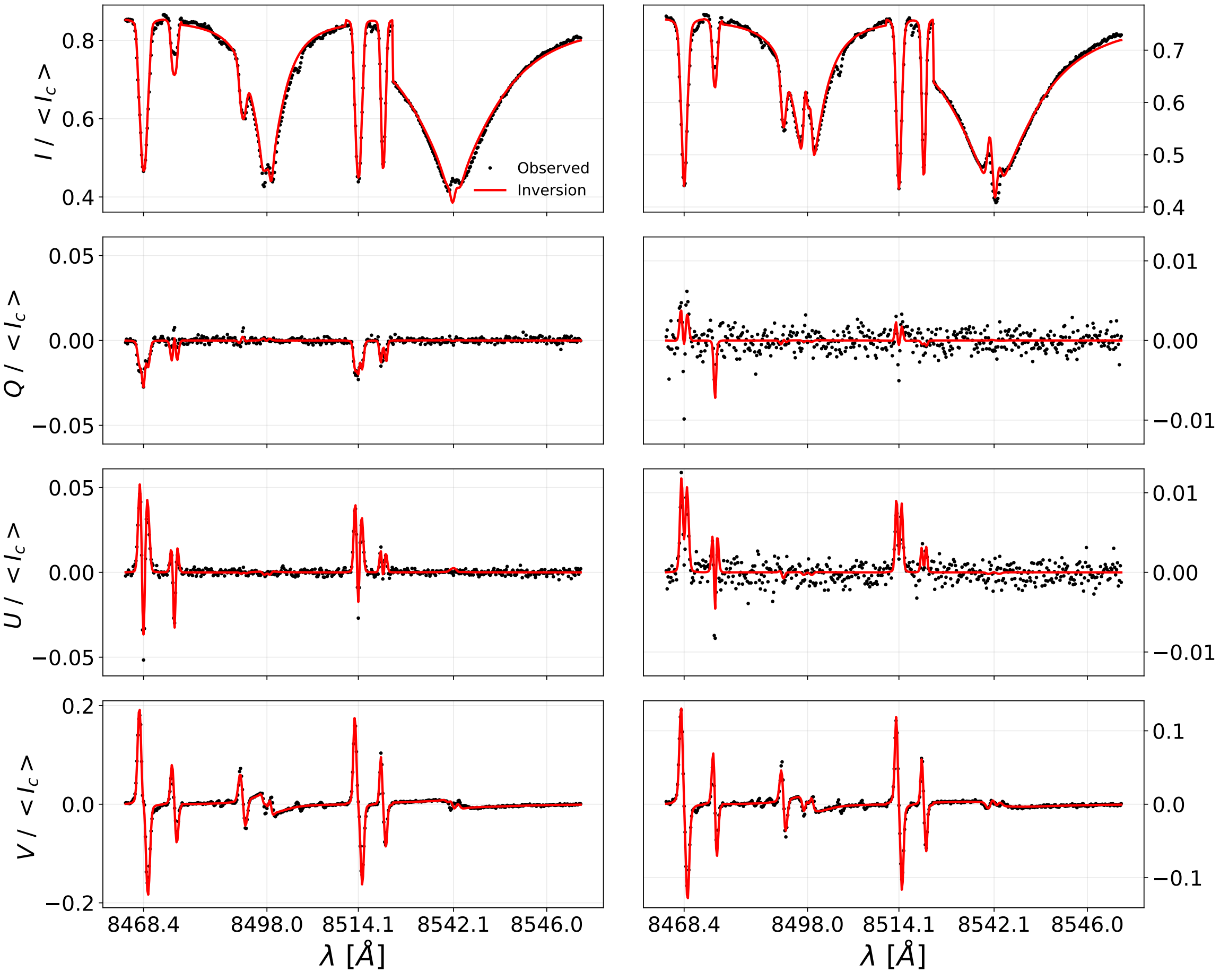

**Figure 2.** Comparison between the observed and inverted Stokes profiles for two representative pixels, one located in Pore 1 (left) and the other in Pore 2 (right), with magnetic-field strengths of 1676 and 861 G at $\log\tau = -2.0$, respectively. From top to bottom, the panels display the normalized Stokes $I$, $Q$, $U$, and $V$ profiles. All Stokes profiles are expressed in units of the mean continuum intensity, $\langle I_c \rangle$. The black dots represent the observed SCIP spectra, whereas the red line shows the best-fit synthetic profile obtained from the simultaneous multiline inversion. The inversion includes the photospheric Fe I lines at 8468, 8471, 8497, 8514, and 8515 Å, together with the chromospheric Ca II lines at 8498 and 8542 Å. The very good agreement between the observed and synthetic profiles demonstrates the ability of the inversion to reproduce simultaneously the photospheric and chromospheric polarization signals over a broad range of magnetic-field strengths.

The procedure was carried out separately for each pore in each of the four SCIP raster maps. For each case, a region enclosing a single pore and excluding opposite polarity magnetic features throughout the sampled atmosphere was first selected. At the reference layer ($\log\tau = -0.5$), an initial magnetic mask was defined by selecting all pixels satisfying $|B_{\rm LoS}| > B_{\rm ref}$, where $B_{\rm LoS}$ denotes the component of the magnetic field along the LoS. Nominal reference thresholds of 900, 850, and 800 G were adopted for Pores 1, 2, and 3, respectively. To quantify the uncertainty associated with the definition of the reference mask, $B_{\rm ref}$ was varied by $\pm 100$ G around these nominal values in steps of 50 G, resulting in five independent realizations for each pore and raster map.

At the reference optical depth layer, the flux-weighted centroid of the reference magnetic mask was determined. Connected magnetic regions were then identified as groups of neighboring pixels sharing a common edge or corner. At each atmospheric layer, only the connected component containing the reference centroid was retained, while disconnected pixels and nearby magnetic structures were discarded.

The magnetic flux enclosed by the reference mask was computed using the absolute value of the LoS magnetic field as

$$\Phi_0 = \sum_i |B_{{\rm LoS},i}|\, A_{\rm pix}, \tag{2}$$

where the sum extends over all pixels belonging to the retained connected component and $A_{\rm pix} = 4.74 \times 10^{13}\ {\rm cm}^2$ is the projected physical area of 1 pixel, computed from the pixel scale of $0\farcs095\ {\rm pixel}^{-1}$ and a conversion factor of $725\ {\rm km\ arcsec}^{-1}$.

At each higher atmospheric layer, the threshold applied to $|B_{\rm LoS}|$ was iteratively adjusted until the enclosed magnetic flux matched the reference value $\Phi_0$. The magnetic cross-sectional area at each optical depth layer was then computed as

$$A(\tau) = N(\tau)\, A_{\rm pix}, \tag{3}$$

where $N(\tau)$ is the number of pixels belonging to the retained connected component. The magnetic expansion factor was defined as

$$f_{\rm exp}(\tau) = \frac{A(\tau)}{A_0}, \tag{4}$$

where $A_0 = A(\log\tau = -0.5)$ is the area enclosed by the reference mask.

As a convenient global measure of the magnetic expansion, we define

$$E = f_{\rm exp}(\log\tau = -4.5), \tag{5}$$

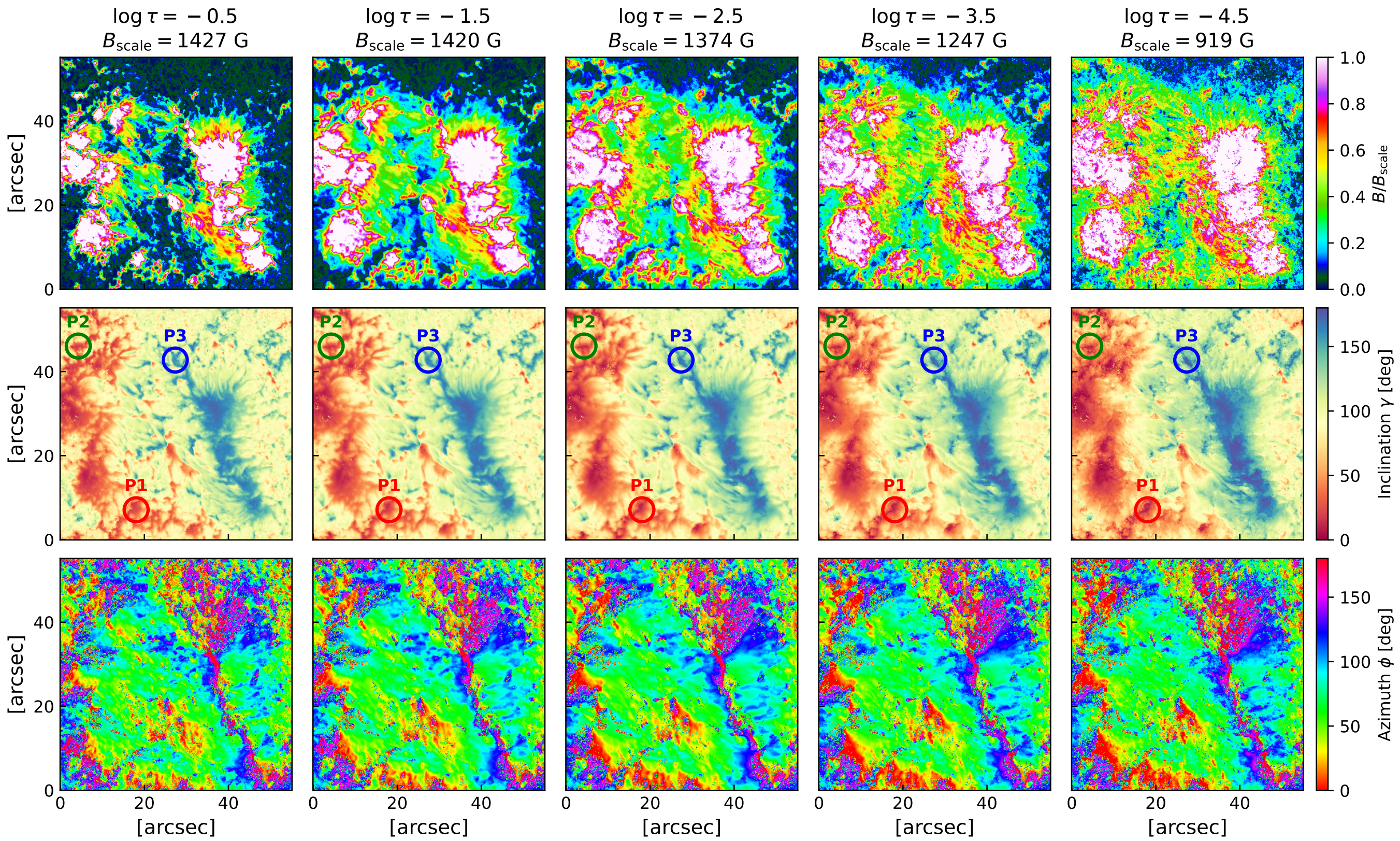

**Figure 3.** Vector magnetic field inferred from the simultaneous multiline inversions. The columns correspond to five equally spaced optical depth layers from $\log\tau = -0.5$ to –4.5. From top to bottom, the panels show the magnetic-field strength ($B$), inclination ($\gamma$), and azimuth ($\phi$), respectively. The magnetic-field strength maps are normalized independently at each optical depth layer by the reference value $B_{\rm scale}$, indicated above each panel. The red, green, and blue circles in the inclination maps indicate Pores 1, 2, and 3, respectively. No 180° azimuth disambiguation has been applied; the azimuth maps correspond directly to the values returned by the inversion. The sequence illustrates the spatially coherent evolution of the inferred magnetic-field vector across the sampled optical depth range.

which corresponds to the total expansion over the optical depth range considered in this work.

The complete analysis was repeated for all adopted reference thresholds, producing a family of expansion curves for each pore and raster map. The mean expansion profile was adopted as the final measurement, while the standard deviation among the different threshold values was taken as the associated uncertainty.

The resulting expansion curves were then parameterized using the empirical relation

$$\log f_{\rm exp} = a + b(-\log\tau) + c(-\log\tau)^2\,, \qquad (6)$$

which provides a compact representation of the measured expansion profiles. The curvature parameter $c$ quantifies the curvature of the expansion law and therefore the acceleration of the magnetic expansion with optical depth. A linear dependence ($c = 0$) corresponds to an expansion proceeding at a constant rate in logarithmic optical depth, whereas positive values of $c$ indicate that the expansion rate increases toward higher atmospheric layers. The constant-flux method implicitly assumes that the magnetic flux enclosed by the reference mask is conserved throughout the atmospheric layers sampled by the inversions. In principle, low-lying magnetic connections near to opposite polarity features could violate this assumption by redirecting part of the magnetic flux into neighboring structures, thereby modifying the inferred flux contours. However, our sensitivity analysis shows that the characteristic increase of the magnetic cross section toward lower optical depths is preserved when the reference magnetic-field threshold is varied, suggesting that the inferred expansion is not critically dependent on the exact definition of the reference mask. We further verified that the inferred expansion is not driven by weak, noise-dominated magnetic signals. Even at $\log\tau = -4.5$, virtually all pixels enclosed by the constant-flux masks show observed Stokes $V$ amplitudes above $5\sigma$ in both Ca II lines.

The resulting measurements allow us to investigate the temporal evolution of the magnetic expansion across the four consecutive SCIP raster scans. The results are presented in the following section.

## 4. Results

### *4.1. Expansion of the Magnetic Field with Height*

Figures 4 and 5 summarize the application of the constant-flux method to the three pores analyzed in this work. Figure 4 presents the magnetic contours enclosing the same magnetic flux at different optical depths, while Figure 5 quantifies the corresponding expansion curves and their temporal evolution.

The continuum maps in Figure 4 show the photospheric intensity structure, while the colored contours trace the magnetic masks obtained from the constant-flux analysis. In all three pores, the contours remain compact in the lower photosphere (between $\log\tau = -0.5$ and $-2.5$) and progressively expand toward lower optical depths. The flux-weighted centroids of the magnetic masks are also shown at each optical depth layer, allowing the displacement of the magnetic structures with height to be directly visualized. The magnetic cross section generally increases toward lower optical depths, although the detailed evolution of the contours varies among pores and raster maps and the expansion is not isotropic. In several cases, the magnetic structures expand preferentially along specific directions, and the orientation of this preferential expansion evolves during the observing sequence. For example, Pore 1 shows a systematic displacement of the flux-weighted centroid with optical depth in all four raster maps, indicating that its magnetic expansion is not purely vertical and is consistent with an inclined magnetic structure. Pores 2 and 3 also show centroid displacements, although these are less systematic across the observing sequence.

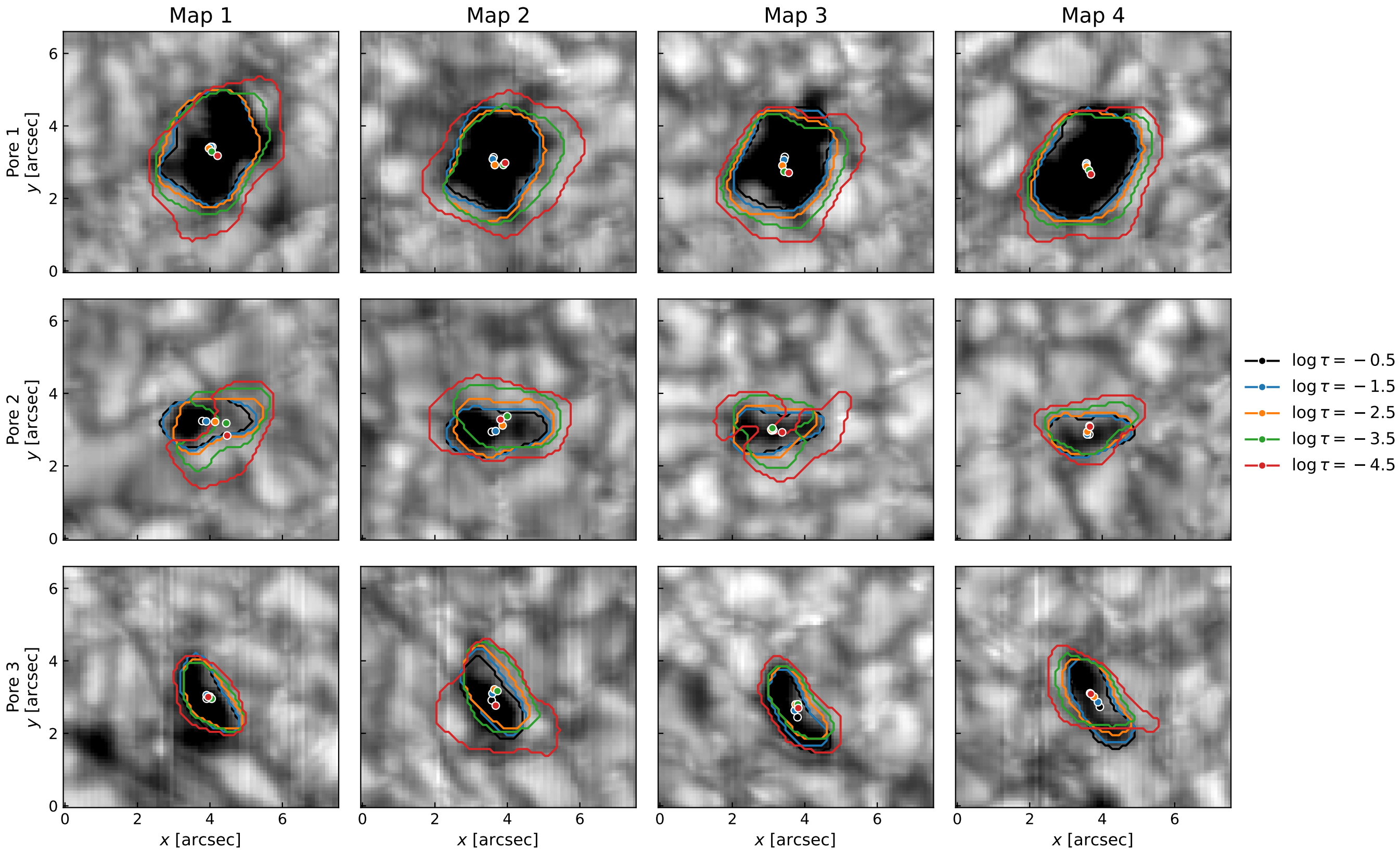

**Figure 4.** Constant magnetic flux contours derived from the multiline inversions for the three pores analyzed in this work. Each row corresponds to a different pore, while the four columns show the four consecutive SCIP raster maps. The background grayscale image displays the Fe I 8468 Å continuum intensity ($\Delta\lambda = -0.555$ Å). Colored contours delineate the boundaries enclosing the same magnetic flux at optical depths $\log\tau = -0.5$, $-1.5$, $-2.5$, $-3.5$, and $-4.5$, shown from black to red. The colored markers the flux-weighted centroid of each magnetic mask, using the same color coding as the corresponding contour. The progressive outward displacement of the constant-flux contours with decreasing optical depth reveals the systematic expansion of the magnetic field with height.

The upper left panel of Figure 5 shows the magnetic expansion factor (Equation (4)) as a function of optical depth for Pore 1 in Map 1, computed using five different reference magnetic-field thresholds to define the initial magnetic mask. The resulting curves are nearly indistinguishable, demonstrating that the inferred expansion is largely insensitive to the choice of the reference magnetic-field threshold within the 800–1000 G range explored here. Considering all four raster maps, the mean expansion factor $E = A(-4.5)/A(-0.5)$ varies by at most 0.6%, 3.9%, and 3.3% within the adopted threshold ranges for Pores 1, 2, and 3, respectively. The largest sensitivity for an individual raster map is found for Pore 2 in Map 3 (16.8%), whereas the remaining maps of this pore vary by less than 3.4%. Importantly, the characteristic increase of the magnetic cross section toward lower optical depths is preserved throughout the tested threshold ranges. The uncertainty bars shown in the remaining panels correspond to the standard deviation obtained from the five reference thresholds considered.

The 12 lower panels of Figure 5 display the expansion curves for the three pores during the four consecutive SCIP raster maps. All three pores show an overall increase of the expansion factor toward lower optical depths. The measured expansion profiles exhibit a relatively modest increase between $\log\tau = -0.5$ and $\approx -2.5$, followed by a substantially steeper rise for $\log\tau \lesssim -3.5$. This behavior indicates that the magnetic expansion becomes more pronounced toward the upper atmospheric layers.

The empirical quadratic function defined by Equation (6) provides a compact parameterization of the measured expansion profiles for all pores and raster maps. The fitted values of the curvature parameter $c$, which quantify the curvature of the empirical expansion profiles, are indicated in each panel, while their temporal evolution is summarized in the upper right panel of Figure 5. The quoted uncertainties correspond to the formal $1\sigma$ errors of the least-squares quadratic fits, derived from the covariance matrix of the fitted parameters. The quadratic representation is intended solely to characterize the overall curvature of the expansion profiles and should not be interpreted as a physical model. Small departures from monotonic expansion are present in some individual profiles. In particular, for Pore 3 the normalized areas inferred directly from the constant-flux masks fall slightly below unity at intermediate optical depths in Maps 1 and 3, with deviations of at most approximately 5%. The polynomial representation can additionally accentuate this behavior. Given their small magnitude, we do not interpret these local deviations as evidence for a physical contraction, but rather as an indication of the uncertainties in the inferred expansion. Importantly, they do not alter the overall increase of the magnetic cross section toward the upper atmospheric layers. The parameter $c$ remains a useful empirical descriptor of how rapidly the expansion accelerates with height. Although small variations are present from map to map, the fitted values of $c$ remain relatively similar throughout the observing sequence. Since the empirical parameterization also includes a linear term, similar values of $c$ do not necessarily imply identical total expansion factors.

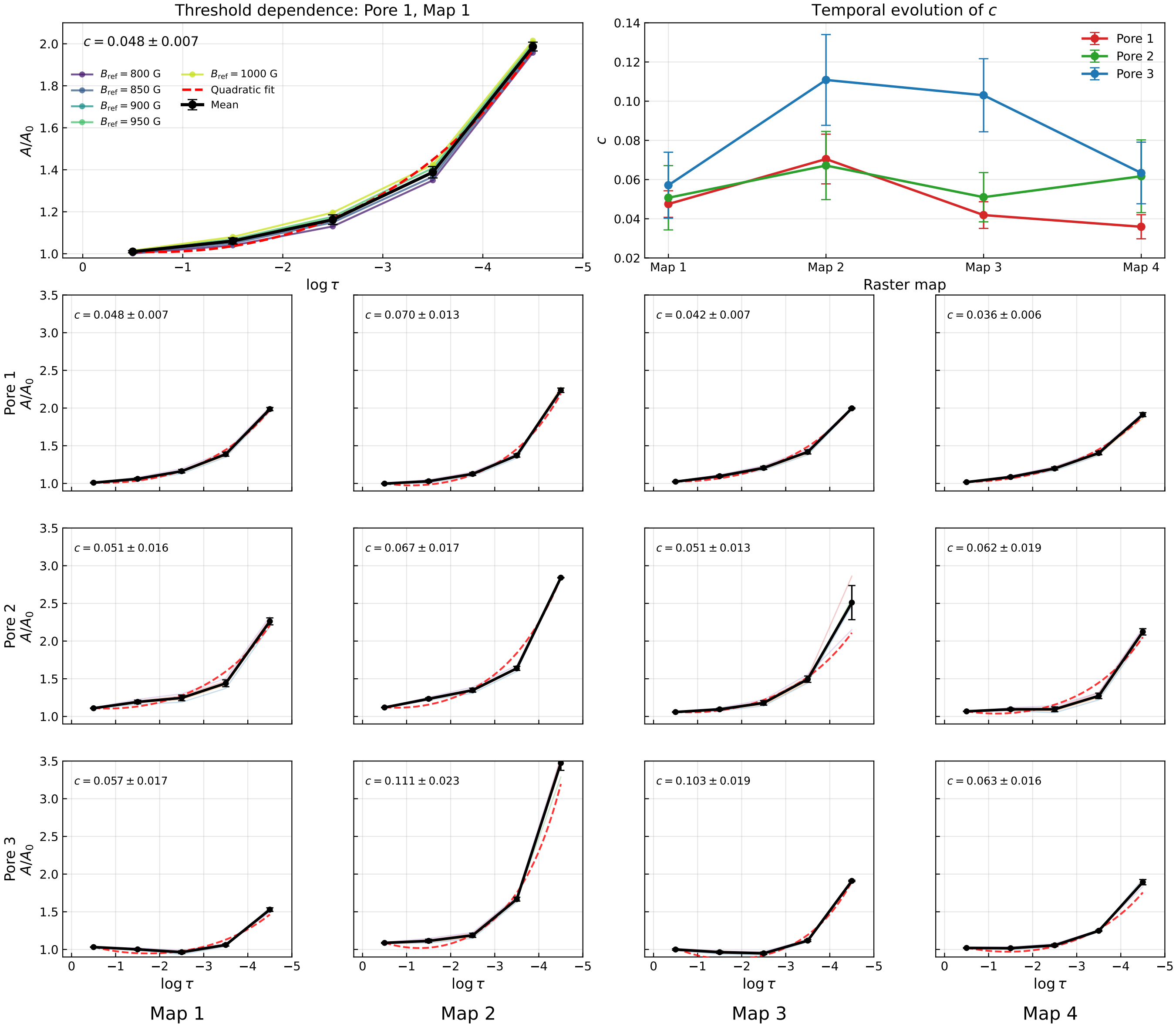

**Figure 5.** Characterization of the magnetic expansion derived from the constant-flux analysis. The upper left panel illustrates the dependence of the expansion profile on the reference magnetic-field threshold used to define the reference magnetic mask at $\log \tau = -0.5$, shown for Pore 1 in Map 1. Colored curves correspond to different reference thresholds, the black curve shows the mean expansion profile, and the red dashed line represents the quadratic fit used to quantify the expansion. The upper right panel displays the temporal evolution of the curvature parameter $c$ (Equation (6)) for the three pores analyzed in this work. The remaining 12 panels show the mean expansion profiles for each pore (rows) and each SCIP raster map (columns), together with their corresponding curvature parameters. Black symbols indicate the measured expansion factor $A/A_0$, while the error bars represent the standard deviation obtained from the threshold analysis. The red dashed curves correspond to the quadratic fits performed in logarithmic space. Larger values of $c$ indicate a more rapid increase of the magnetic area with height.

Averaging over the four raster maps, we obtain total expansion factors of $E = 2.01 \pm 0.09$ for Pore 1, $E = 2.27 \pm 0.13$ for Pore 2, and $E = 2.11 \pm 0.37$ for Pore 3. Thus, all three pores approximately double their magnetic cross section between $\log \tau = -0.5$ and –4.5, corresponding to the lower photosphere and the highest atmospheric layers reliably sampled by the multiline inversions. The corresponding mean values of the curvature parameter are $c = 0.046 \pm 0.011$, $0.055 \pm 0.017$, and $0.079 \pm 0.025$ for Pores 1, 2, and 3, respectively. Together, the contour evolution

**Table 1**
Mean Physical and Magnetic Properties of the Three Pores Analyzed in This Work

| Pore | $R_{eq}$ (arcsec) | $\Delta T$ (K) | $B_{core}$ (G) | $\langle B_{env} \rangle$ (G) | $\Phi_0$ ($10^{19}$ Mx) | $E$ | $c$ |
|---|---|---|---|---|---|---|---|
| 1 | 1.371 | −391 ± 54 | 1704 ± 44 | 603 ± 45 | 3.84 | 2.01 ± 0.09 | 0.046 ± 0.011 |
| 2 | 0.797 | −139 ± 45 | 1684 ± 76 | 361 ± 32 | 1.10 | 2.27 ± 0.13 | 0.055 ± 0.017 |
| 3 | 0.818 | −180 ± 46 | 1639 ± 37 | 488 ± 52 | 1.05 | 2.11 ± 0.37 | 0.079 ± 0.025 |

**Note.** Values correspond to averages over the four SCIP raster maps. The equivalent radius $R_{eq}$ is defined by Equation (7), the magnetic flux $\Phi_0$ by Equation (2), the total expansion factor $E$ by Equation (5), and the curvature parameter $c$ by Equation (6). The magnetic quantities and the temperature contrast are evaluated at $\log \tau = -0.5$. The core magnetic field $B_{core}$ is defined as the median of the upper 10% of the $|B_{LoS}|$ values within the reference magnetic mask. The temperature contrast is defined as $\Delta T = T_{pore} - T_{env}$, while $\langle B_{env} \rangle$ is the mean magnetic-field strength measured in a 10 pixel wide annulus surrounding the reference magnetic mask. Uncertainties in the physical quantities correspond to the temporal standard deviation of the four raster maps. The uncertainty in $E$ is the standard error of the mean, while the uncertainty in $c$ combines the temporal standard deviation with the mean formal uncertainty of the quadratic fits.

and the expansion curves indicate that the height-dependent magnetic expansion remains broadly stable throughout the observing sequence.

### *4.2. Expansion Dependence on Pore Properties*

To investigate whether the magnetic expansion depends on the intrinsic properties of the magnetic structures, we computed, for each raster map, the reference magnetic area, $A_0$, the equivalent radius

$$R_{eq} = \sqrt{\frac{A_0}{\pi}}, \tag{7}$$

the magnetic flux enclosed by the reference mask, $\Phi_0$, the temperature contrast between the pore and its immediate surroundings, $\Delta T = T_{pore} - T_{env}$, the core magnetic-field strength, $B_{core}$, and the mean magnetic-field strength of the surrounding atmosphere, $\langle B_{env} \rangle$, all evaluated at $\log \tau = -0.5$. The core magnetic field is defined as the median of the upper 10% of the $|B_{LoS}|$ values within the reference magnetic mask, while the surrounding magnetic field is measured in a 10 pixel wide annulus immediately outside the reference mask. These quantities were averaged over the four SCIP raster maps to obtain representative values for each pore. The resulting physical properties, together with the total expansion factor $E \equiv f_{exp}(\log \tau = -4.5)$ and the curvature parameter $c$, are summarized in Table 1.

The three pores span equivalent radii between approximately $0\overset{''}{.}8$ and $1\overset{''}{.}4$, magnetic fluxes between $1 \times 10^{19}$ and $4 \times 10^{19}$ Mx, and core magnetic-field strengths close to 1.7 kG. Their small spatial scales illustrate the need for high-spatial-resolution and stable spectropolarimetric observations to characterize the expansion of magnetic structures through the solar atmosphere.

Although the three pores display remarkably similar core magnetic-field strengths, they exhibit different expansion factors and curvature parameters, suggesting that the strength of the magnetic core alone is insufficient to account for the observed differences in expansion. The pores also differ substantially in the thermodynamic and magnetic properties of their immediate surroundings. However, when the four raster maps are examined individually, no consistent relationship is found between the expansion and either the temperature deficit or the surrounding magnetic-field strength. For example, while the map of Pore 1 with the largest temperature deficit exhibits the weakest expansion, the opposite behavior is found for Pores 2 and 3. Therefore, given the small number of magnetic structures analyzed here and their temporal variability, our observations do not allow us to establish which physical properties control the differences in expansion. A larger statistical sample will be required to determine whether the thermodynamic and magnetic properties of the surrounding atmosphere play a systematic role.

Finally, the total expansion factor, $E \equiv f_{exp}(\log \tau = -4.5)$, and the curvature parameter $c$ provide complementary diagnostics of the magnetic geometry of the pores. The former quantifies the overall increase of the magnetic area between the reference layer and the highest atmospheric layer sampled by the inversions, whereas the latter describes how rapidly the expansion itself accelerates with optical depth. The differences in the relative values of $E$ and $c$ among the three pores illustrate that these two quantities characterize complementary aspects of the three-dimensional magnetic topology.

## 5. Discussion and Conclusions

The expansion of magnetic concentrations with height is a fundamental consequence of magnetic flux conservation and lateral pressure balance in a stratified atmosphere (H. C. Spruit 1976; S. K. Solanki 1993). Classical flux tube models predict that magnetic structures remain relatively compact and nearly vertical in the deep photosphere while progressively broadening into canopy-like configurations toward the chromosphere (H. C. Spruit 1976; S. K. Solanki 1993; S. K. Solanki et al. 1999). Our observations are qualitatively consistent with this picture. The constant-flux contours shown in Figure 4 reveal an overall increase of the magnetic area with decreasing optical depth, while the magnetic-field inclination maps in Figure 3 show progressively more transverse fields surrounding the pores. The expansion is modest in the lower photosphere (between $\log \tau = -0.5$ and –2.5) and becomes significantly stronger at lower optical depths ($\log \tau \lesssim -2.5$). This height-dependent expansion provides an observational constraint on a fundamental geometric property involved in the formation of magnetic canopies.

A key aspect of our analysis is that the expansion is measured directly from the inferred magnetic field without assuming an idealized flux tube geometry. Classical models generally describe circular or axisymmetric magnetic structures (H. C. Spruit 1976; S. K. Solanki 1993), whereas the observed pores exhibit irregular morphologies and evolve continuously with time. By following the magnetic structure enclosing a fixed magnetic flux, the constant-flux method provides an observational definition of magnetic expansion that can be applied directly to realistic solar magnetic structures. To the best of our knowledge, this work presents the first direct measurements of the height-dependent magnetic expansion of solar pores.

Compared with previous observational studies, which investigated canopy formation through Stokes profile asymmetries (M. J. Martínez González et al. 2012), center to limb variations (A. Pietarila et al. 2010), spectropolarimetric inversions based on a more limited set of spectral lines (D. Buehler et al. 2015; C. Kuckein 2019), or recent three-dimensional magnetic-field reconstructions from coordinated CLASP2.1 and Hinode observations (R. Ishikawa et al. 2025), the simultaneous multiline inversions presented here provide direct access to the three-dimensional magnetic-field stratification over a substantially broader range of optical depths. The very good agreement between observed and synthetic Stokes profiles, together with the spatial coherence of the inferred atmospheric parameters and the magnetic response function analysis, supports the use of simultaneous multiline inversions to constrain the magnetic-field stratification over a broad range of optical depths from the photosphere to the lower chromosphere.

The constant-flux analysis further provides two complementary quantities to characterize the expansion of the magnetic field. The total expansion factor $E$ measures the overall increase of the magnetic cross-sectional area between the lowest and highest atmospheric layers sampled by the inversions, whereas the curvature parameter $c$ quantifies how rapidly the expansion rate increases with height, thereby characterizing the acceleration of the expansion. Two magnetic structures may therefore exhibit similar total expansion factors while displaying different vertical expansion profiles. We find that all three pores approximately double their magnetic cross section between $\log \tau = -0.5$ and –4.5, while the expansion curves remain broadly stable throughout the 48 minute observing sequence. The robustness of the expansion against variations in the initial magnetic mask indicates that the inferred height-dependent expansion is not critically dependent on the precise choice of the reference threshold. Previous observational studies have also reported magnetic expansion in a variety of solar magnetic structures. For example, C. Kuckein (2019) found area increases of 7%–51% in isolated bright points by comparing magnetic maps inferred from Si I inversions at $\log \tau = -1.0$ and $-2.5$, whereas I. Juanikorena Berasategi et al. (2026) measured photosphere to chromosphere expansion factors of 1.7–2.1 in active region plage using Fe I and Ca II diagnostics. In contrast, the simultaneous multiline inversions and constant-flux method employed here recover the height dependence of the magnetic cross-sectional area over the atmospheric range sampled by the inversions, revealing that the expansion is not uniform but becomes significantly stronger toward the upper atmospheric layers.

The comparison among the three pores also shows that differences in magnetic expansion cannot be straightforwardly related to a single property of the magnetic structures. Although all three pores display remarkably similar core magnetic-field strengths ($B_{\rm core} \approx 1.65$–1.70 kG), they exhibit different expansion behaviors and are embedded in thermodynamically and magnetically different environments. However, the relationships suggested by the temporally averaged pore properties are not consistently reproduced in the individual raster maps. Given the small number of magnetic structures analyzed here and their temporal variability, our observations therefore do not allow us to establish a robust dependence of the expansion on either the properties of the magnetic core or those of its surroundings. Theoretical models indicate that the expansion of magnetic flux concentrations may depend on the combined effects of their intrinsic magnetic and thermodynamic structure and their interaction with the surrounding atmosphere (S. K. Solanki & O. Steiner 1990). Larger observational samples will be required to determine the relative importance of these effects.

In summary, this work demonstrates the potential of simultaneous multiline inversions for studying the three-dimensional magnetic topology of solar magnetic structures. Combined with the constant-flux analysis, these inversions provide a quantitative framework for measuring the LoS expansion of magnetic fields without assuming simplified geometries. Our comparison of the three analyzed pores shows that the observed differences in magnetic expansion cannot be robustly associated with any single property of the magnetic structures or their surroundings. Larger observational samples will be required to determine which physical parameters govern these differences. The methodology can be applied to larger samples of pores, network elements, and sunspots, as well as to future observations obtained with facilities such as DKIST (T. R. Rimmele et al. 2020) and the European Solar Telescope (EST; C. Quintero Noda et al. 2022), providing new observational constraints for theoretical flux tube models, radiative magnetohydrodynamic simulations of magnetic canopy formation, and solar wind models based on magnetic flux tube expansion (e.g., Y.-M. Wang & N. R. Sheeley 1990; C. N. Arge & V. J. Pizzo 2000).

## Acknowledgments

J.C.T.A. acknowledges financial support from grant PID2024-156066OB-C55 and PID2024-156538NB-I00, both funded by MCIN/AEI/10.13039/501100011033 and by “ERDF A way of making Europe” S.J.G.M. acknowledges grant RYC2022-037565-I and CK grant RYC2022-037660-I, both funded by MCIN/AEI/10.13039/501100011033 and by “ESF Investing in your future.” The authors wish to acknowledge the contribution of the IAC High-Performance Computing support team and hardware facilities to the results of this research. Sunrise III is supported by funding from the Max-Planck-Förderstiftung (Max Planck Foundation); NASA under grants #80NSSC18K0934 and #80NSSC24M0024 (“Heliophysics Low Cost Access to Space” program); and the ISAS/JAXA Small Mission-of-Opportunity program and JSPS KAKENHI grant Nos. JP18H05234 and JP23K25916. This research has received financial support from the European Union’s Horizon 2020 research and innovation program under grant agreement No. 824135 (SOLARNET) and No. 101097844 (WINSUN) from the European Research Council (ERC). It has also been funded by the Deutsches Zentrum für Luft- und Raumfahrt e.V. (DLR; grant No. 50 OO 1608). The Spanish contributions have been funded by the Spanish MCIN/AEI under projects RTI2018-096886-B-C5, and PID2021-125325OB-C5, and from “Center of Excellence Severo Ochoa” awards to IAA-CSIC (SEV-2017-0709 and CEX2021-001131-S), all cofunded by European REDEF funds, “A way of making Europe.” This work is part of grant CEX2025-001609-S, awarded to the Instituto de Astrofísica de Canarias under the Severo Ochoa Centre of Excellence program and funded by MICIU/AEI/10.13039/501100011033. We acknowledge ChatGPT for AI-assisted language editing.

## ORCID iDs

Juan Carlos Trelles Arjona https://orcid.org/0000-0001-9857-2573
María Jesús Martínez González https://orcid.org/0000-0001-5560-7502
Christoph Kuckein https://orcid.org/0000-0002-3242-1497
Sergio Javier González Manrique https://orcid.org/0000-0002-6546-5955
Manuel Collados https://orcid.org/0000-0002-6210-9648
Sami K. Solanki https://orcid.org/0000-0002-3418-8449
Andreas Lagg https://orcid.org/0000-0003-1459-7074
Achim Gandorfer https://orcid.org/0000-0002-9972-9840
Jose Carlos del Toro Iniesta https://orcid.org/0000-0002-3387-026X
Yukio Katsukawa https://orcid.org/0000-0002-5054-8782
Pietro Bernasconi https://orcid.org/0000-0002-0787-8954
Alex Feller https://orcid.org/0009-0009-4425-599X
Tino L. Riethmüller https://orcid.org/0000-0001-6317-4380
Alberto Álvarez-Herrero https://orcid.org/0000-0001-9228-3412
Masahito Kubo https://orcid.org/0000-0001-5616-2808
H. N. Smitha https://orcid.org/0000-0003-3490-6532
David Orozco Suárez https://orcid.org/0000-0001-8829-1938
Valentín Martínez Pillet https://orcid.org/0000-0001-7764-6895
Francisco Javier Bailén https://orcid.org/0000-0002-7318-3536
Julian Blanco Rodríguez https://orcid.org/0000-0002-2055-441X
Juan Sebastián Castellanos Durán https://orcid.org/0000-0003-4319-2009
Edvarda Harnes https://orcid.org/0009-0002-6808-5154
Johannes Hölken https://orcid.org/0000-0001-6029-7529
Francisco A. Iglesias https://orcid.org/0000-0003-1409-1145
Ryohtaroh T. Ishikawa https://orcid.org/0000-0002-4669-5376
Yusuke Kawabata https://orcid.org/0000-0001-7452-0656
Takuma Matsumoto https://orcid.org/0000-0002-1043-9944
Takayoshi Oba https://orcid.org/0000-0002-7044-6281
Azaymi L. Siu-Tapia https://orcid.org/0000-0003-0175-6232
Hanna Strecker https://orcid.org/0000-0003-1483-4535
Dušan Vukadinovic https://orcid.org/0000-0003-1971-5551
Hirohisa Hara https://orcid.org/0000-0001-5686-3081
Toshifumi Shimizu https://orcid.org/0000-0003-4764-6856